\documentclass[%
 reprint,
 amsmath,amssymb,
 aps,
prd,
]{revtex4-2}

\usepackage{graphicx}
\usepackage{dcolumn}
\usepackage{bm}
\usepackage[colorlinks,urlcolor=blue,linkcolor=blue,anchorcolor=blue,citecolor=blue]{hyperref}

\begin{document}

\preprint{APS/123-QED}

\title{Double-sphere enhanced optomechanical spectroscopy constrains symmetron dark energy}

\author{Jiawei Li}
\author{Ka-Di Zhu}
\email{zhukadi@sjtu.edu.cn}
\affiliation{Key Laboratory of Artificial Structures and Quantum Control (Ministry of Education),\\
School of Physics and Astronomy, Shanghai Jiao Tong University, 800 DongChuan Road,
Shanghai 200240, China
}

\date{\today}

\begin{abstract}
Screened scalar fields such as the symmetron provide a viable description of dark energy yet their laboratory detection remains challenging. We propose an optomechanical scheme to constrain symmetron interactions using two optically levitated nanospheres inside a cavity. The symmetron-mediated interaction induces an effective coupling which leads to a measurable splitting in the optomechanical resonance spectrum. We forecast constraints in the regime $\mu \sim 10^{-2}\text{eV}-10^{-4}\text{eV}$, which shows that this approach can improve existing laboratory bounds by up to several orders of magnitude, demonstrating the sensitivity of optomechanical spectroscopy to screened fifth forces. 
\end{abstract}

\maketitle


\section{\label{sec1}Introduction}

The accelerating expansion of universe has been confirmed by a variety of observation results including cosmic microwave background measurements~\cite{chang2022snowmass2021cosmicfrontiercosmic,Balbi_2001}, type Ia supernovae~~\cite{ruiter2024typeiasupernovaprogenitors,blondin2025typeiasupernovae}, and analyses of large-scale structure~~\cite{doi:10.1142/S0217751X22500865,sarkar2024impactlargescalestructureformation}. These observations are well described within the framework of the $\Lambda$ cold dark matter ($\Lambda$CDM) cosmological model~~\cite{Brax_2018,Riess_1998,Schmidt_1998,ostriker1995cosmicconcordance,doi:10.1126/science.284.5419.1481}.  However, it is disconcerting that there exists a discrepancy of about 60 orders of magnitude between the observed vacuum energy density, $\rho_{\Lambda}^{\mathrm{obs}} \sim 10^{-120} M_{\mathrm{Pl}}^{4}$, and its theoretical expectation from quantum zero-point fluctuations, $\rho_{\Lambda}^{\mathrm{theo}} \sim 10^{-60} M_{\mathrm{Pl}}^{4}$ with cutoff $\Lambda\sim1\text{TeV}$~\cite{JOYCE20151,LOMBRISER2019134804,Velten_2014,Solà_2013,padilla2015lecturescosmologicalconstantproblem,RevModPhys.61.1}. In addition, the apparent coincidence between the present vacuum energy density and the matter density also remains open problems, known as the coincidence problem~\cite{PhysRevD.78.021302,Gonçalves2005}.

The smallness of the cosmological constant has often been interpreted from an anthropic perspective. In this view, a universe with a much larger vacuum energy would be unable to support the formation of large-scale structures and, consequently, the emergence of life~~\cite{PhysRevLett.59.2607}. While this anthropic argument provides a plausible selection effect, it does not offer a dynamical or physical explanation for the origin of cosmic acceleration, nor does it yield a predictive framework for the underlying structure of spacetime~~\cite{RaphaelBousso_2000,susskind2003anthropiclandscapestringtheory,PhysRevD.16.1762,PhysRevD.27.2848,PhysRevLett.74.846}. 

Cosmic acceleration can also arise from the potential energy of a slowly rolling scalar field, a scenario commonly referred to as quintessence~\cite{Tsujikawa_2013,doi:10.1142/S0217732308027631,PhysRevLett.95.141301}.  However, such models are strongly constrained by local tests of gravity and the equivalence principle~\cite{ValerioFaraoni_2011}. Moreover, the extremely weak coupling to matter required to satisfy these constraints often reintroduces fine-tuning problems, thereby limiting the theoretical appeal of these scenarios~\cite{HERTZBERG2019134878,galaxies10020050}.

An alternative approach is provided by scalar--tensor theories of gravity, which originate from early works by Jordan, Brans and Dicke, and Bergmann~~\cite{Jordan1955-JORSUW,PhysRev.124.925,Bergmann:1968ve}. As modifications of general relativity, these theories allow scalar fields to couple to matter and generically predict the existence of a so-called fifth force with observable consequences on cosmological scales~~\cite{JOYCE20151,Brax_2013}. In high-density environments, however, such effects can be strongly suppressed, ensuring consistency with precision tests of general relativity in the solar system and in laboratory experiments~~\cite{PhysRevLett.93.171104,khoury2010theories,PhysRevD.84.103521}. The way to hide scalar fields from local laboratory detection, named as screening mechanism, which can be broadly classified based on the relation to the local field value $\phi$. The first class of screening mechanisms is governed by a potential $V(\phi)$, including the symmetron~\cite{PhysRevD.84.103521}, chameleon~\cite{PhysRevD.69.044026}, and dilaton~\cite{PhysRevD.84.123504}. And the second class is based on $\partial \phi$ to shut off the fifth force, such as K-mouflage~\cite{PhysRevD.90.023507} and P(x) models~\cite{PhilippeBrax_2013}. Finally, the last class considering $\partial^{2}\phi$ involves the Vainshtein~\cite{VAINSHTEIN1972393} and Galileon theories~\cite{PhysRevD.79.064036}.

In this work we focus on the symmetron model, which can be regarded as a concrete realization of the Damour-Polyakov mechanisms~\cite{PhysRevD.72.043535,PhysRevD.86.044015,DAMOUR1994532}. To be specific, it relies on a $\mathbb{Z}_2$-symmetric action of the symmetry-breaking form, under which whether the $\mathbb{Z}_2$ symmetry is spontaneously broken depends on the local  matter density the scalar field $\phi$ couples to. In high density environment, the effective field potential has only one minimum and remains zero. While in the region with density lower than a critical value, the symmetry is spontaneously broken and $\phi$ acquires a vacuum expectation value $\phi =\pm \mu/\sqrt{\lambda}$ ~\cite{Burrage_2016,Brax_2013,PhysRevD.84.103521}.  Existing constraints already exclude sizable regions of parameter space by variety of experiment design in both cosmical scales and tabletop experiments, including Casimir-force detection~\cite{PhysRevD.75.077101,PhysRevLett.78.5,PhysRevD.101.064065},  quantum nonlocality tests~\cite{y314-4x4s}, astronomy detection ~\cite{RevModPhys.93.015003,PhysRevD.101.083524,PhysRevD.111.084020}, torsion pendulum experiments~\cite{PhysRevLett.98.021101,PhysRevLett.110.031301,PhysRevLett.129.141101}, neutron tests~\cite{article,article1} and atomic interferometry~\cite{Burrage_2015,PhysRevD.94.104069,Burrage_2016,article2}. 

The purpose of the present work is to establish constraints by utilizing a detection strategy based on optomechanical systems. In contrast to conventional force-measurement techniques, our approach exploits the modification of mechanical resonance frequencies induced by the symmetron-mediated interaction.  The proposed scheme provides a stronger constraints by up to $1\sim4$ orders of magnitude across a broad region of the symmetron parameter space within the region of $\mu$ in $10^{-2}$eV $\sim10^{-4}$ eV, complementing existing laboratory and astrophysical bounds. This paper is organized as follows. In Sec.~\ref{2.1}, we briefly review the symmetron mechanism and derive the scalar field profile and fifth force generated by a spherically symmetric source mass. In Sec.~\ref{2.2}, we introduce the optomechanical system considered in this work and show how the symmetron-induced interaction modifies the effective Hamiltonian of the coupled optical and mechanical modes. In Sec.~\ref{3.1}, we discuss the choice of simulation experimental parameters and estimate the influence of environmental effects to our hypothetic experiment. In Sec.~\ref{3.2}, we present the resulting constraints on the symmetron parameter space. Finally, Sec.~\ref{4} summarizes our results and outlines prospects for future improvements.

\section{Model and Theory}\label{2}
\subsection{The symmetron screening mechanism}\label{2.1} Generally speaking, the action of a scalar--tensor theory  can be written in the form ~\cite{PhysRevD.47.5329}
\begin{align}
S =\,& \int d^{4}x \sqrt{-g}
\left(\frac{M_{\mathrm{Pl}}^{2}}{2} R
- \frac{1}{2} (\partial \phi)^{2}
- V(\phi)\right) \nonumber\\
& + S_{\mathrm{matter}}\!\left[A^{2}(\phi) g_{\mu\nu},\psi\right],
\label{eq:2.1.1}
\end{align}
in which $R$ is the Ricci scalar associated with the Einstein-frame metric $g_{\mu\nu}$, and $V(\phi)$ is the scalar-field potential. Matter fields $\psi$ are  coupled to the Jordan-frame metric $ \tilde{g}_{\mu\nu} = A^{2}(\phi)\, g_{\mu\nu}$ which introduces an effective coupling between matter and the scalar field through the conformal factor $A(\phi)$.

Considering a non-relativistic matter source, we note that the equation of motion for $\phi$ derived from Eq.~(\ref{eq:2.1.1}) is
\begin{equation}
\square \phi =\frac{dV_{\mathrm{eff}}(\phi)}{d\phi},
\label{eq:2.1.2}
\end{equation}
where $V_{\mathrm{eff}}(\phi) = V(\phi) + A(\phi)\rho$. Note that the scalar field can be dynamically suppressed in regions of high density by suitably choosing the potential and coupling to matter, by which the screening mechanism works. 

For the symmetron, the simplest incarnation of this mechanism with a $\mathbb{Z}_2$-symmetric action of the symmetry-breaking form is ~\cite{PhysRevD.84.103521,PhysRevLett.104.231301}
\begin{align}
S = {} &
\int d^{4}x \sqrt{-g}
\left(
\frac{M_{\mathrm{Pl}}^{2}}{2} R
- \frac{1}{2} (\partial \phi)^{2}
+ \frac{\mu^{2}}{2} \phi^{2}
- \frac{\lambda}{4} \phi^{4}
\right)
\nonumber \\
& + S_{\mathrm{matter}}
\!\left[
\left( 1 + \frac{\phi^{2}}{2M^{2}} \right)^{2}
g_{\mu\nu}
\right],
\label{eq:2.1.3}
\end{align}
 with the potential
\begin{equation}
V(\phi)
= - \frac{\mu^{2}}{2} \phi^{2}
+ \frac{\lambda}{4} \phi^{4},
\label{eq:2.1.4}
\end{equation}
and the conformal factor given by
\begin{equation}
A(\phi)
= 1 + \frac{\phi^{2}}{2M^{2}}
+ \mathcal{O}\!\left( \frac{\phi^{4}}{M^{4}} \right),
\label{eq:2.1.5}
\end{equation}
in which the model is governed by two parameters $\mu$ and M in mass scales as well as another dimensionless coupling parameter $\lambda$. Here, we drop the high order term since the parameter $M$ is some high mass scale that ($\phi \ll M$). And then it is clear that the effective potential in Eq.~(\ref{eq:2.1.2}) felt by $\phi$ in the non-relativistic scenario is given by
\begin{equation}
V_{\mathrm{eff}}(\phi)
=
\frac{1}{2}
\left(
\frac{\rho}{M^{2}} - \mu^{2}
\right)
\phi^{2}
+ \frac{\lambda}{4} \phi^{4}.
\label{eq:2.1.6}
\end{equation}

Moreover, in vacuum, we consider a spherically symmetric, static, homogeneous, pressureless source of radius $R$ ($\tilde{T} \simeq -\tilde{\rho}$). The equation of motion Eq.~(\ref{eq:2.1.2}) describing the associated scalar field in spherical coordinates reduces to
\begin{equation}
\frac{d^{2}\phi}{dr^{2}}
+ \frac{2}{r}\frac{d\phi}{dr}
=\frac{dV_{\mathrm{eff}}(\phi)}{d\phi}
\label{eq:2.1.7}
\end{equation}

The boundary conditions require that the solution be smooth at the origin and approach its vacuum expectation value $\phi_0=\mu/\sqrt{\lambda}$ at infinity,
\begin{equation}
\frac{d\phi}{dr}(0) = 0,
\qquad
\phi(r \to \infty) = \phi_{0}.
\label{eq:2.1.8}
\end{equation}

With the above boundary conditions, the interior and exterior solutions of Eq.(\ref{eq:2.1.7}) are given by
\begin{equation}
\phi(r)=\frac{B}{r}\sinh\!\left(\sqrt{\alpha}\,\frac{r}{R}\right),\qquad r < R.
\label{eq:2.1.9}
\end{equation}
\begin{equation}
\phi(r)=\phi_{0}+\frac{C}{r}\exp\!\left(-\sqrt{2}\,\mu r\right),\qquad r > R,
\label{eq:2.1.10}
\end{equation}
where $\alpha=\rho R^2/M^2$ is introduced as the thin-shell factor characterizing the screening strength, and $B$ and $C$ are two undetermined constants.

The coefficients $B$ and $C$ are determined by imposing the continuity of the scalar field and its derivative at the interface $r=R$. The resulting expressions are
\begin{equation}
B=\phi_{0} R\frac{\sinh\!\left(\sqrt{\alpha}\right)-\sqrt{\alpha}\cosh\!\left(\sqrt{\alpha}\right)}{\sqrt{\alpha}\cosh\!\left(\sqrt{\alpha}\right)+\sqrt{2}\,\mu R \sinh\!\left(\sqrt{\alpha}\right)}
\label{eq:2.1.11}
\end{equation}
and
\begin{equation}
C=\phi_{0} R\frac{1 + \sqrt{2}\,\mu R}{\sqrt{\alpha}\cosh\!\left(\sqrt{\alpha}\right)+\sqrt{2}\,\mu R \sinh\!\left(\sqrt{\alpha}\right)}.
\label{eq:2.1.12}
\end{equation}

In this work. we focus on the symmetron-induced fifth force in the thin-shell limit ($\alpha \gg 1$),
Eq.(\ref{eq:2.1.10}) reduces to
\begin{equation}
\phi\simeq\phi_{0}-\phi_{0}\left(1 - \frac{1}{\sqrt{\alpha}}\right)\frac{R}{r}\exp\!\left[-\sqrt{2}\,\mu (r - R)\right].
\label{eq:2.1.13}
\end{equation}
For $\mu r \ll 1$, the exterior solution in the far field region $(R\ll r)$ for two dense spheres further simplifies to
\begin{equation}
\phi\simeq\phi_{0}-\phi_{0}\frac{R}{r}.
\label{eq:2.1.14}
\end{equation}

Considering Eq.~(\ref{eq:2.1.5}),  the acceleration induced by the symmetron fifth force on a test object can be expressed in terms of the scalar field as
\begin{equation}
{\mathbf{a}}=-\frac{d \ln A(\phi)}{d\phi}\, {\nabla}\phi\approx-\frac{1}{M^2}\phi\nabla\phi \qquad \phi\ll M .
\label{eq:2.1.15}
\end{equation}
Substituting Eq.~(\ref{eq:2.1.14}) into Eq.~(\ref{eq:2.1.15}), one can derive the fifth force sourced from a strongly screened  sphere to a test object with mass $m_{\text{test}}$ by:
\begin{equation}
F_s=\frac{m_{\text{test}}\phi_0^2R}{M^2r^2}.
\label{eq:2.1.16}
\end{equation}

We now return to Eq.~(\ref{eq:2.1.10}) and consider the thick-shell limit $\alpha \ll 1$. In this case, the scalar field profile can be approximated as
\begin{equation}
\phi\simeq\phi_{0}-\phi_{0}\,\frac{\alpha}{3}\frac{R}{r}\exp\!\left[-\sqrt{2}\,\mu (r - R)\right],
\label{eq:2.1.17}
\end{equation}
which further simplifies in the regime $\mu r \ll 1$ to
\begin{equation}
\phi\simeq\phi_{0}-\phi_{0}\,\frac{\alpha}{3}\frac{R}{r}.
\label{eq:2.1.18}
\end{equation}

Substituting Eq.~(\ref{eq:2.1.18}) into Eq.~(\ref{eq:2.1.15}), the fifth force of non-screened source with $m_{\text{source}}$ is:
\begin{equation}
F_{\text{ns}}=\frac{m_{\text{test}}m_{\text{source}}\phi_0^2}{4\pi M^4r^2}.
\label{eq:2.1.19}
\end{equation}

Comparing Eqs.~(\ref{eq:2.1.16}) and~(\ref{eq:2.1.19}), one finds that the fifth force is suppressed by a factor $\lambda_{\text{sphere}}=3M^2/\rho R^2$. Physically, this reflects the fact that the fifth force is no longer proportional to the total mass of the sphere, but only to a fraction $\lambda_{\text{sphere}} \ll 1$. Additionally, The source and test objects are strongly screened equally in this work. Consequently, for a system consisting of two strongly screened spheres, the interaction fifth force takes the form\cite{PhysRevD.101.064065}
\begin{equation}
F=F_{\text{ns}}\times\lambda_{\text{source}}\lambda_{\text{test}}=\frac{4\pi R_{\text{source}}R_{\text{test}}\phi_0^2}{r^2}.
\label{eq:2.1.20}
\end{equation}

\subsection{The cavity optomechanical system}\label{2.2}
\begin{figure}[h]
\centering
\includegraphics[width=0.93\linewidth]{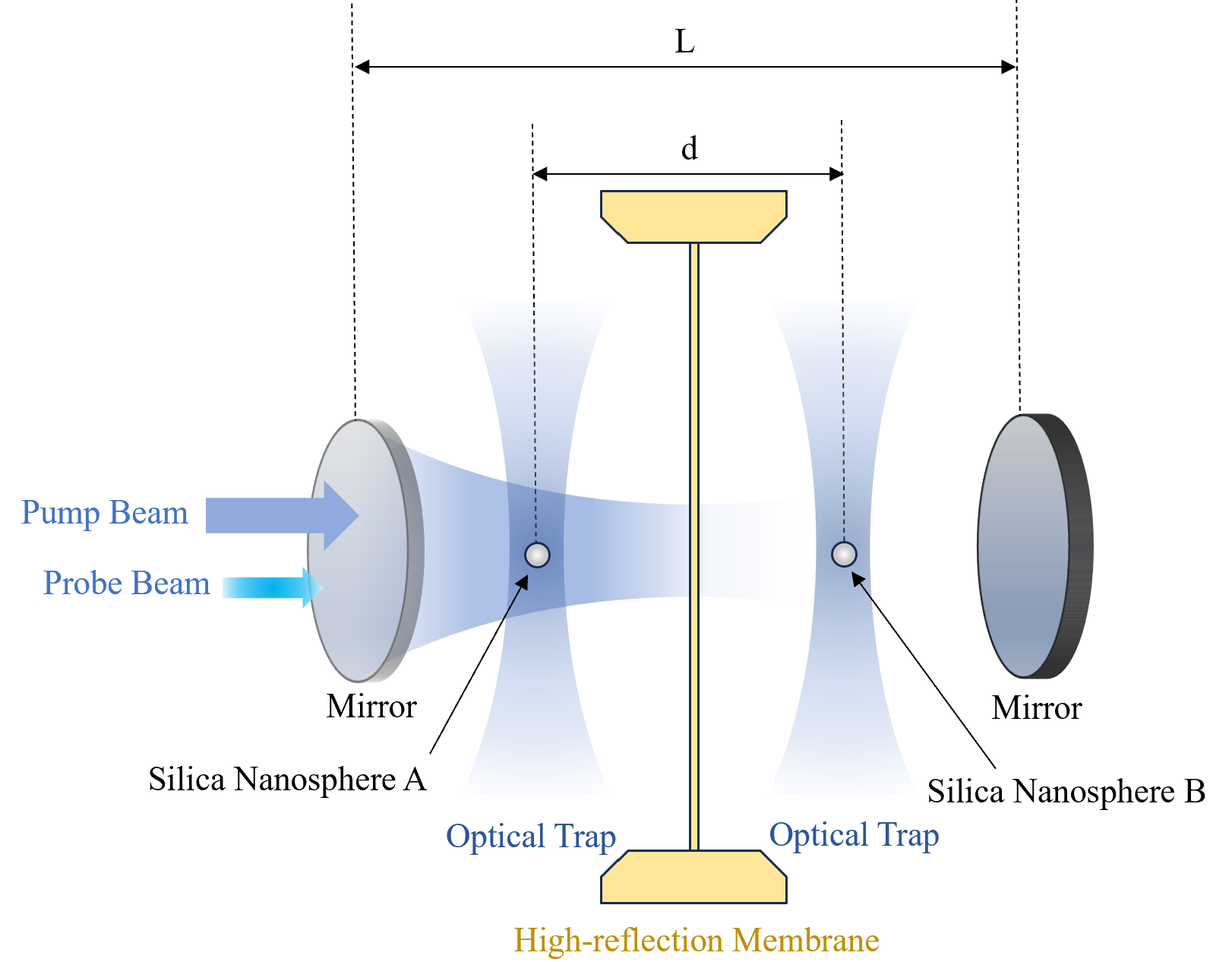}
\caption{\label{fig:1}Schematic diagram of the experimental setup. The system consists of two optically trapped silica nanospheres (A and B) with the same radius $R=0.5\mu$m, which are confined inside a Fabry--Pérot cavity formed by two mirrors separated by a distance $L$. A strong pump beam together with a weak probe beam is injected through the left mirror to drive and probe the cavity field. These two nanospheres are localized at the opposite sides of a high-reflection membrane placed at the cavity center. The distance between the two nanospheres is denoted by $d=4\mu$m. And the membrane also serves as a reflective mirror that isolates the two trapping regions.}
\end{figure}

Our scheme consider a system shown in Fig.~\ref{fig:1}. Two uniform fused silica spheres A and B are optically trapped at the center of the cavity with intrinsic vibration frequency $\omega_a$ and $\omega_b$. The two spheres are separated by a gold-coated SiC membrane, which serves to suppress electrostatic and Casimir background forces by preventing direct electromagnetic coupling between the spheres~~\cite{PhysRevD.78.022002,PhysRevLett.90.151101}. The membrane, with a thickness of $\sim 10$nm is expected to be sufficiently stiff as reported in the previous works ~\cite{Nguyen2017SuperiorRU,inproceedings,doi:10.1021/acsphotonics.3c00968}. The left mirror, the sphere A and the high-reflection membrane construct a new cavity, and we only consider the optomechanical interaction Hamiltonian between the light and the sphere A in the following part. We utilized a pump light with frequency $\omega_p $ and a relatively weak probe light with frequency $ \omega_{\text{pr}}$~\cite{PhysRevD.106.095007,PhysRevLett.105.101101,PhysRevD.95.044014}. 

The free Hamiltonian of the system are given as follows: $H_{\text{free}}=\hbar \omega_a \hat{o}_a^\dagger\hat{o}_a+\hbar \omega_b \hat{o}_b^\dagger \hat{o}_b+\hbar \omega_{cav} \hat{a}^\dagger \hat{a}$, consisting of the free Hamiltonian of sphere A, B and the cavity mode $\omega_{cav}$. Here $\hat{a}$, $\hat{o}_a$ and $\hat{o}_b$ are the annihilation operators of the cavity mode and the two mechanical modes, respectively. In a frame rotating at the pump frequency, the total Hamiltonian of the system
takes the form~\cite{PhysRevA.77.033804,PhysRevA.63.023812}
\begin{align}
H = {} &
\hbar \Delta_{\mathrm{pu}} \hat{a}^\dagger \hat{a}+ \hbar \omega_a \hat{o}_a^\dagger \hat{o}_a+ \hbar \omega_b \hat{o}_b^\dagger \hat{o}_b+ \hbar g \hat{a}^\dagger \hat{a}\left(\hat{o}_a^\dagger + \hat{o}_a\right)\nonumber \\ & - i \hbar \Omega_p\left(\hat{a} - \hat{a}^\dagger\right)- i \hbar \Omega_{\mathrm{pr}}\left(\hat{a} e^{i\delta t}- \hat{a}^\dagger e^{-i\delta t}\right)+ H_{\mathrm{int}}.
\label{eq:2.2.1}
\end{align}
Here $\Delta_{\mathrm{pu}} = \omega_{cav} - \omega_p$ is the pump–cavity detuning and $\delta = \omega_{\mathrm{pr}} - \omega_p$ is the pump–probe detuning. And $\Omega_p=\sqrt{2 P_p \kappa/\hbar\omega_{\mathrm{p}}},\Omega_{\mathrm{pr}}=\sqrt{2 P_{\mathrm{pr}} \kappa/\hbar\omega_{\mathrm{pr}}}$ are the Rabi frequencies of the pump and probe fields, respectively. The optomechanical interaction between the dielectric nanosphere A and the cavity field is described by the coupling Hamiltonian $\hbar g \hat{a}^\dagger \hat{a}(\hat{o}_a^\dagger + \hat{o}_a)$, where $g$ is the coupling strength. The last term $H_{\mathrm{int}}$ accounts for the symmetron-induced interaction potential energy between the two levitated spheres A and B.

Considering Eq.~(\ref{eq:2.1.20}), the resulting potential energy is given by
\begin{equation}
V(r)=-\frac{4\pi R_{\text{source}}R_{\text{test}}\phi_0^2}{r}.
\label{eq:2.2.2}
\end{equation}
Here the distance between the two spheres can be written as $r = d + x_a - x_b$,where $x_{a,b}$ stand for the displacements of the spheres A and B from their equilibrium positions. Expanding Eq.~(\ref{eq:2.2.2}) to second order in the small displacements $|x_a|,|x_b|\ll d$, one obtains
\begin{equation}
V(r)\simeq - 4\pi R_{\text{source}}R_{\text{test}}\phi_0^2 \\  \Bigg[\frac{1}{d}- \frac{x_a - x_b}{d^{2}}+ \frac{(x_a - x_b)^2}{d^{3}}\cdots\Bigg]
\label{eq:2.2.3}
\end{equation}

For Eq.~(\ref{eq:2.2.3}), One can find the first term remain constant, while the second term corresponds to a steady force that does not contribute to the interaction dynamics~\cite{articlejl}. The third term, with second order of $x_a-x_b$, describes a coupling between the motions of the two spheres. Considering only the lowest-order coupling term, the interaction Hamiltonian can be obtained by quantization of the mechanical degrees of freedom of spheres A and B. With the rotating-wave approximation, we have 
\begin{align}
H_{\mathrm{int}}=- 8\pi R_{\text{source}}R_{\text{test}}\phi_0^2x_ax_b \nonumber \\ \simeq\hbar \Omega\left(\hat{o}_a^\dagger \hat{o}_b+\hat{o}_a \hat{o}^\dagger_b\right),
\label{eq:2.2.4}
\end{align}
where the coupling strength is given by
\begin{eqnarray}
\Omega=&&-\frac{8\pi R_{\text{source}}R_{\text{test}}\phi_0^2}{\hbar d^3}\sqrt{\frac{\hbar}{2m_a \omega_a}}\sqrt{\frac{\hbar}{2m_b \omega_b}} \\= &&-\frac{4\pi R_{\text{source}}R_{\text{test}}\mu^2}{\lambda d^3}\sqrt{\frac{1}{m_am_b \omega_a \omega_b}}
\label{eq:2.2.5}
\end{eqnarray}
The coefficient $\Omega$ is defined as the strength of the symmetron-induced interaction between the two spheres, with the dimension of frequency.

Then we define the operators $s_i=\hat{o}_i^\dagger + \hat{o}_i\quad (i=a,b)$, which describe the position quadratures of the mechanical oscillators. By considering the commutation relations that $[\hat{o}_i,\hat{o}_i\dagger] = 1$ and $[a,a^\dagger] = 1$, substituting Eqs.~(\ref{eq:2.2.1}) and (\ref{eq:2.2.5}) into the Heisenberg equations of motion, we have the mean field evolution of $s_i$ and $a$ described by the quantum Langevin equations with additional damping terms:
\begin{align}
\frac{d \langle \hat{a} \rangle}{dt}
&=
-\left(i\Delta_{\mathrm{pu}} + \kappa \right)\langle \hat{a} \rangle
+ i g \langle s_a \rangle \langle \hat{a} \rangle
+ \Omega_p
+ \Omega_{\mathrm{pr}} e^{-i\delta t},\nonumber
\\ & \quad+ \sqrt{\kappa_{ex}} \langle\hat{a}_{in} \rangle + \sqrt{\kappa_0} \langle\hat{f}_{in} \rangle,
\label{eq:2.2.6}
\\
\frac{d^{2}\langle s_a \rangle}{dt^{2}}
&+
\gamma_1 \frac{d\langle s_a \rangle}{dt} +(\omega_a^{2} + \Omega^{2})\langle s_a \rangle-\Omega(\omega_a + \omega_b)\langle s_b \rangle \nonumber
\\ & =2g \omega_a \langle \hat{a}^\dagger \rangle \langle \hat{a} \rangle + \langle \hat{\xi}\rangle ,
\label{eq:2.2.7}
\\
\frac{d^{2}\langle s_b \rangle}{dt^{2}}
&+
\gamma_2 \frac{d\langle s_b \rangle}{dt}
+
(\omega_b^{2} + \Omega^{2})\langle s_b \rangle
-
\Omega(\omega_a + \omega_b)\langle s_a \rangle \nonumber
\\ &=-2 g \Omega \langle \hat{a}^\dagger \rangle \langle \hat{a} \rangle +\langle \hat{\xi}\rangle  .
\label{eq:2.2.8}
\end{align}

Here $\kappa=\kappa_0+\kappa_{\mathrm{ex}}$ denotes the total cavity decay rate, and $\gamma_i$ ($i=a,b$) are the damping rates of the mechanical modes of the spheres. The Langevin noise operator $\hat{a}_{\text{in}}(t)$ and $\hat{f}_{\text{in}}$  represent the quantum and thermal noise operators associated with the optical and mechanical dissipation channels, respectively. With correlation function written as $\langle \hat{O}(t) \hat{O}(t_0) \rangle \sim \delta(t - t_0)$, these operators are supposed to have the mean value $\langle \hat{a}_{\text{in}}(t) \rangle=\langle \hat{f}_{\text{in}}(t) \rangle = 0$. The operator $\hat{\xi}$ accounts for the influence of the thermal bath arising from non-Markovian stochastic processes and Brownian motion \cite{PhysRevA.63.023812,gardiner2004quantum}. It also has a zero mean value with correlation function is given by $\left\langle \hat{\xi(t)^{\dagger}}\hat{\xi(t')} \right\rangle = \gamma\omega\int1/2\pi\omega e^{-i\omega(t-t')} \left[ 1 + \coth\left( \hbar\omega/2k_bT \right) \right]d\omega$.  

To solve Eqs.(\ref{eq:2.2.6})-(\ref{eq:2.2.8}), we introduce the following ansatz~\cite{10.5555/1817101}:
\begin{align}
\langle \hat{a}(t) \rangle
&=
a_0 + a_{+} e^{-i\delta t} + a_{-} e^{i\delta t},
\label{eq:2.2.9}
\\
\langle s_a(t) \rangle
&=
s_{a0} + s_{a+} e^{-i\delta t} + s_{a-} e^{i\delta t},
\label{eq:2.2.10}
\\
\langle s_b(t) \rangle
&=
s_{b0} + s_{b+} e^{-i\delta t} + s_{b-} e^{i\delta t}.
\label{eq:2.2.11}
\end{align}

Taking Eqs.\eqref{eq:2.2.9}–\eqref{eq:2.2.11} into Eqs.\eqref{eq:2.2.6}-\eqref{eq:2.2.8} and dropping high order terms with weak probe $\Omega_{\mathrm{pr}}$, one can obtain the following equations with the same time dependence:
\begin{align}
s_{a+}=&
\frac{
\Omega(\omega_a + \omega_b) s_{b+}
+ 2 g \omega_a (a_0^{*} a_{+} + a_0 a_{-}^{*})
}
{-\delta^{2} - i\gamma_1 \delta + \omega_a^{2} + \Omega^{2}},
\label{eq:2.2.12}
\\
s_{a-}=&
\frac{
\Omega(\omega_a + \omega_b) s_{b-}
+ 2 g \omega_a (a_0^{*} a_{-} + a_0 a_{+}^{*})
}
{-\delta^{2} + i\gamma_1 \delta + \omega_a^{2} + \Omega^{2}},
\label{eq:2.2.13}
\\
s_{b+}=&
\frac{
\Omega(\omega_a + \omega_b) s_{a+}
- 2 g \Omega (a_0^{*} a_{+} + a_0 a_{-}^{*})
}
{-\delta^{2} - i\gamma_2 \delta + \omega_b^{2} + \Omega^{2}},
\label{eq:2.2.14_1}
\\
s_{b-}=&
\frac{
\Omega(\omega_a + \omega_b) s_{a-}
- 2 g \Omega (a_0^{*} a_{-} + a_0 a_{+}^{*})
}
{-\delta^{2} + i\gamma_2 \delta + \omega_b^{2} + \Omega^{2}},
\label{eq:2.2.14}
\\
a_{+}=&
\frac{\Omega_{\mathrm{pr}} + i a_0 g s_{a+}}
{i\Delta_{\mathrm{pu}} + \kappa - i\delta - i g s_{a0}},
\label{eq:2.2.15}
\\
a_{-}=&
\frac{i a_0 g s_{a-}}
{i\Delta_{\mathrm{pu}} + \kappa + i\delta - i g s_{a0}},
\label{eq:2.2.16}
\\
a_0=&
\frac{\Omega_p}
{i\Delta_{\mathrm{pu}} + \kappa - i g s_{a0}},
\label{eq:2.2.17}
\\
(\omega_a^{2} +& \Omega^{2}) s_{a0}
-
\Omega(\omega_a + \omega_b) s_{b0}=2g \omega_a |a_0|^{2},
\label{eq:2.2.18}
\\
(\omega_b^{2} +& \Omega^{2}) s_{b0}
-
\Omega(\omega_a + \omega_b) s_{a0}=
-2 g \Omega |a_0|^{2},
\label{eq:2.2.19}
\end{align}
with the last three equations are the steady state equations which are time independent.

Solving the above equations one can obtain 
\begin{align}
a_{+}=&\frac{\Omega_{\mathrm{pr}}\left[ Z(W - X) + U N_0 \right]}{Z(W^{2} - X^{2}) + 2 X U N_0},
\label{eq:2.2.20}
\end{align}
with the coefficient defined as follows: $W = \kappa - i\delta,X = i\Delta_{\mathrm{pu}} - i g s_{a0}, Y_i = -\delta^{2} - i\gamma_i \delta + \omega_i^{2} + \Omega^{2}\quad(i = 1,2), Z= Y_1 Y_2 - \Omega^{2}(\omega_a + \omega_b)^{2}, U= 2 i g^{2} \omega_a Y_2 - 2 i g^{2} \Omega^{2}(\omega_a + \omega_b)$. And here the cavity photon number $N_0 = |a_0|^{2}$ is obtained by solving $\Omega_p^{2}=\left[\kappa^{2}+\left(\Delta_{\mathrm{pu}} - g s_{a0}\right)^{2}\right]N_0$.

The final step of the optical calculation part is to calculate the transmission of the probe field. With the help of the input-output relation 
\begin{align}
a_{\text{out}}(t) &= a_{\text{in}}(t) - \sqrt{2\kappa} a(t)\nonumber\\
&=( \Omega_p/ \sqrt{2\kappa}e^{-i\omega_pt} +\Omega_s/\sqrt{2\kappa}e^{-i(\omega_p+\delta)t})  \nonumber \\
&-(\sqrt{2\kappa}a_0e^{-i\omega_pt}+\sqrt{2\kappa}a_+e^{-i(\omega_p+\delta)t} \nonumber \\
&+\sqrt{2\kappa}a_-e^{-i(\omega_p-\delta)t}).
\label{eq:2.2.21}
\end{align}
where $a_{\text{in}}$ and $a_{\text{out}}$ are the input drives and output fields operators.

Concentrate on the terms with frequency $ \omega_{\text{pr}}=\omega_p +\delta$, we derive the transmission of the output probe field component, defined as the ratio of the output field and input field at the probe frequency~\cite{bowen2015quantum}:

\begin{eqnarray}
\mathcal{T}(\omega_s) = \frac{\Omega_s /\sqrt{2\kappa}-\sqrt{2\kappa}a_+}{\Omega_s/ \sqrt{2\kappa}}. 
\label{eq:2.2.22}   
\end{eqnarray}
\section{NUMERICAL RESULTS}\label{3}
\subsection{The detection of the frequency shift}\label{3.1}
This section aims to provide practical parameters and to demonstrate numerical results about the cavity transmission spectrum with noise analysis. For the optical system setup, we consider a cylindrical cavity of length $L=31\,\mathrm{mm}$ and mirror radius $R_c=10\,\mathrm{mm}$, with the radius of curvature of the left mirror set to $R_L=89\,\mathrm{mm}$. The cavity finesse is assumed to be $F=30$, which corresponds to a decay rate $\kappa=\pi c/2FL=4.98\times10^8\,\mathrm{Hz}$. The cavity resonance frequency is taken to be $\omega_{\mathrm{cav}}=\omega_p$. For a pump beam with wavelength $\lambda=1064\,\mathrm{nm}$ and power $P_p=1.9\mu$W, the waist of the intracavity optical field and the corresponding Rabi frequency are estimated to be $w=0.12\,\mathrm{mm}$ and $\Omega_p\approx10^{11}\,\mathrm{Hz}$, respectively~\cite{9452539}. On the mechanical side, the silica spheres A and B is set to have the same mass and volume that $m_a=m_b=1.2\times10^{-15}$kg, $V_a=V_b=5.24\times10^{-19}$m$^3$, and also to oscillate at the same intrinsic frequency  $\omega_a=\omega_b=1.23\times10^6$Hz~\cite{article3}. And the trapping laser beam injected into the cavity is also chosen to have the wavelength $\lambda_\text{{laser}}=1064$nm and work at the power P=0.17W ~\cite{Deli2020}. 

Based on the above parameters, several derived quantities can now be evaluated. The optomechanical coupling strength $g\approx1.4$Hz can be derived from ~\cite{doi:10.1073/pnas.0912969107}
\begin{eqnarray}
g=\frac{3}{4}\frac{V_{sphere}}{V_c}\frac{\epsilon-1}{\epsilon+2}\omega_{cav}.
\label{eq:3.1.1}   
\end{eqnarray}
Here $\epsilon=3.75$ is the relative permittivity for silica with $V_c=\pi L w^2/4$ is the cavity mode volume. We further estimate the air-molecule-induced mechanical damping rate of the spheres to be $\gamma_i\approx1.5\times10^{-7}$Hz($i=a,b$) with 
\begin{equation}
\gamma_i=\frac{3P}{R_i\rho_iv}
\label{eq:3.1.2}
\end{equation}
where P=$10^{-10}$torr represents is the residual gas pressure inside the cavity and $v=\sqrt{k_BT/m_{gas}}$ is the mean velocity of air molecules at temperature T=200K~\cite{doi:10.1073/pnas.0912969107,Hunger_2010}.  

\begin{figure*}
\includegraphics[width=0.93\textwidth]{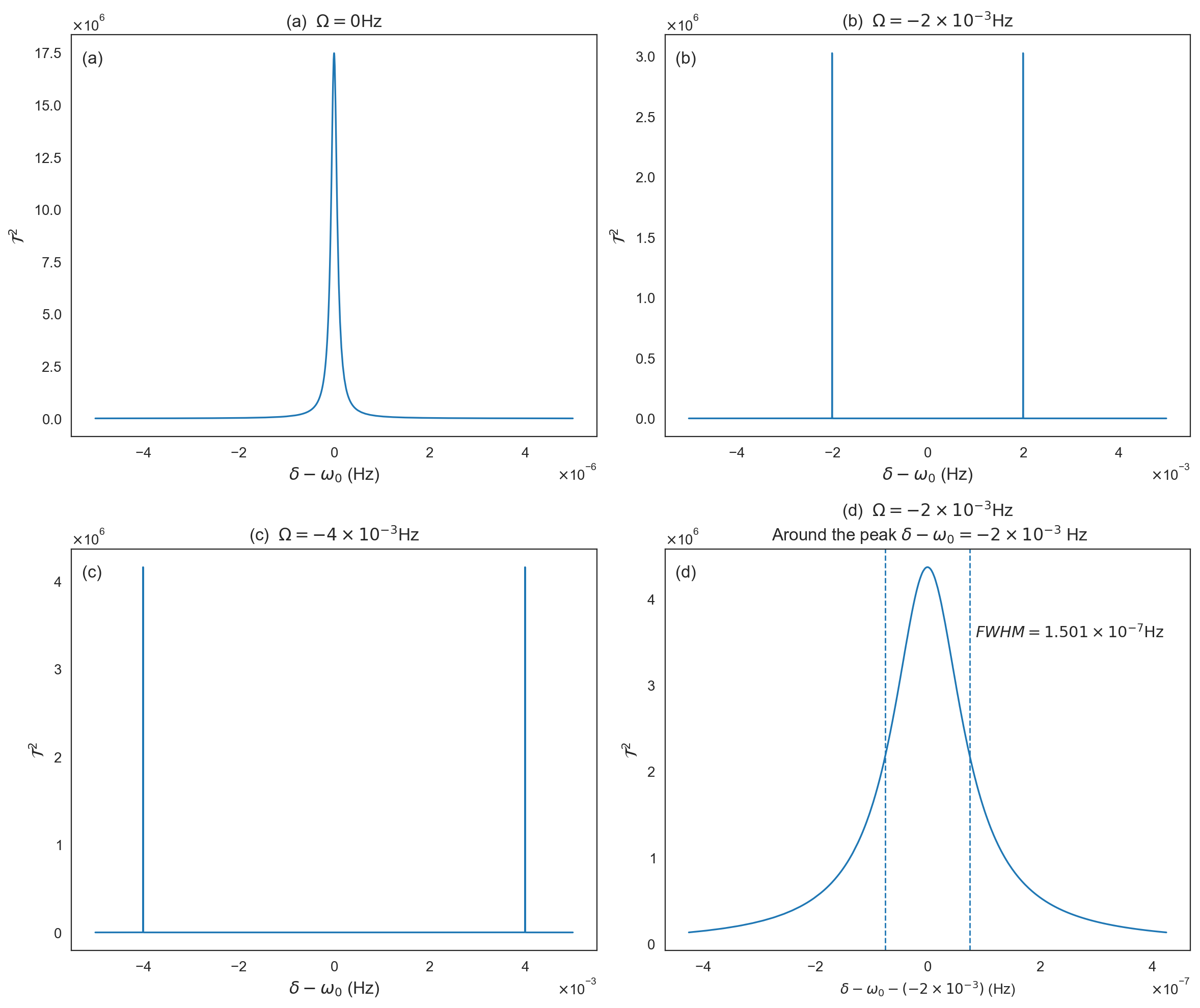}
\caption{\label{fig:2}Transmission $\mathcal{T}^2$ of the probe field as a function of $\delta-\omega_0$ for different values of the effective coupling parameter $\Omega$. (a) In the absence of the symmetron-induced interaction ($\Omega=0$), a single sharp resonance peak appears at $\delta-\omega_0=0$. (b) The original resonance splits into two distinct peaks located symmetrically around the central frequency for $\Omega=-2\times10^{-3}\mathrm{Hz}$. (c) The situation for $\Omega=-4\times10^{-3}\mathrm{Hz}$.(d) A magnified view of the resonance around $\delta-\omega_0=-2\times10^{-3}\,\mathrm{Hz}$ for $\Omega=-2\times10^{-3}\mathrm{Hz}$, where the full width at half maximum (FWHM) is extracted as $1.501\times10^{-7}\,\mathrm{Hz}$, setting the achievable frequency resolution of the system.}
\end{figure*}

Now we proceed to simulate the probe spectrum through Eq.~\eqref{eq:2.2.22} with all parameters determined. The results in $\mathcal{T}^2$ to $\delta-\omega_0$ spectra are shown in Fig.~\ref{fig:2}, where $\omega_0=\omega_{a,b}=1.23\times10^6$Hz is the intrinsic frequency of sphere A and B. Firstly, in Fig.~\ref{fig:2}(a), we first consider the trivial case without fifth-force interaction ($\Omega=0$), in which a single enhanced resonance peak appears at $\delta-\omega_0=0$. 

Moreover, in Fig.~\ref{fig:2}(b) and Fig.~\ref{fig:2}(c), we present the spectral response when the symmetron-induced coupling between spheres A and B is included, with $\Omega=-2\times10^{-3}\mathrm{Hz}$ and $-4\times10^{-3}\mathrm{Hz}$, respectively. To explain the phenomena, we introduce $\lvert N \rangle$, $\lvert n_1 \rangle$, and $\lvert n_2 \rangle$ as the number states of the cavity photons, and the phonons of spheres A and B, respectively. In the absence of the fifth-force interaction ($\Omega=0$), it can been found that The transition from $\lvert N+1, n_1, n_2 \rangle $ to $ \lvert N, n_1+1, n_2 \rangle$ corresponds to the single eigenmode with eigenfrequency $\omega_0$. When the Hamiltonian is modified by the fifth-force term in Eq.~\eqref{eq:2.2.4}, the eigen states of the system are instead replaced by the symmetric and antisymmetric hybrid states $\lvert \pm\rangle$ of $\lvert N, n_1 + 1, n_2 \rangle$ and $\lvert N,n_1, n_2 + 1 \rangle$. These two hybrid states give rise to two new mechanical eigenmodes which correspond to eigen frequencies shifts by $\pm \Omega$ from the uncoupled mechanical frequency $\omega_0$. Thus, as shown in Fig.~\ref{fig:2}(b) and Fig.~\ref{fig:2}(c), the symmetron-induced coupling strength $\Omega$ has a simple relation to the frequency splitting $\omega_{sp}$ that 
\begin{eqnarray}
\omega_{sp}=2\Omega
\label{eq:3.1.2-1}
\end{eqnarray}

The frequency resolution of the system is determined by the full width at half maximum (FWHM) of the resonance peak, which is calculated to be $\omega_{\mathrm{FWHM}}\approx1.501\times 10^{-7}\mathrm{Hz}$ in Fig.~\ref{fig:2}(d). Therefore, the minimum resolvable peak splitting is estimated as $\omega_{sp}=1.501\times 10^{-7}\mathrm{Hz}$.

The next step is to analyze the system-level noise, which mainly consists of electromagnetic noise and thermomechanical noise. To characterize the electromagnetic contribution, we restrict our attention to the left optical cavity and consider only the coupling between sphere A and the shield membrane. Under this configuration, the system can be effectively described by a cavity optomechanical model in which the levitated microsphere and the membrane interact via an electrostatic dipole-dipole coupling. The corresponding interaction potential takes the form
\begin{equation}
U = - \frac{\mu_a \mu_S}{2\pi \varepsilon_0 r'^{3}},
\label{eq:3.1.3}
\end{equation}
where $\mu_a$ and $\mu_S$ denote the electric dipole moments of the microsphere and the shield membrane, respectively. The instantaneous separation is expressed as $r' = d' + x_a - x_S$, where $x_S$ represents the membrane displacement from equilibrium and $d'\approx2~\mu\mathrm{m}$ is the equilibrium distance between sphere A and the membrane surface~\cite{articlejl}.  Following the same procedure in Eqs.~\eqref{eq:2.2.2}-\eqref{eq:2.2.5}, we perform a Taylor expansion around $x_a-x_S\approx0$, from which the dipole-interaction Hamiltonian can be obtained as
\begin{equation}
H_E \approx - \hbar \Psi \left( \hat{o}_a^\dagger \hat{o}_S + \hat{o}_a \hat{o}_S^\dagger \right),
\label{eq:3.1.4}
\end{equation}
with the coupling strength
\begin{equation}
\Psi=\frac{3\mu_i \mu_s}{\pi \varepsilon_0 d'^{5}}\frac{1}{\sqrt{M_a M_S \omega_a \omega_S}}.
\label{eq:3.1.5}
\end{equation}
Here $\hat{o}_S^\dagger$ ($\hat{o}_S$) are the bosonic creation (annihilation) operators. The membrane is assumed to have  mass $M_s\approx6\times10^{-8}$kg and mechanical frequency $\omega_S=70$Mhz for a typical 1 cm-radius Au-coated SiC membrane~\cite{10.1063/1.4866268}.While the dipole moments are given by $\mu_a \approx 1.6\times10^{-23}C\cdot\mathrm{m}$ and $\mu_S \approx1.6\times 10^{-18}C\cdot\mathrm{m}$~\cite{PhysRevLett.117.101101}. With the above parameters, we can calculate the resulting coupling strength as $\Psi\approx0.8$Hz.

\begin{figure}[h]
\centering
\includegraphics[width=0.93\linewidth]{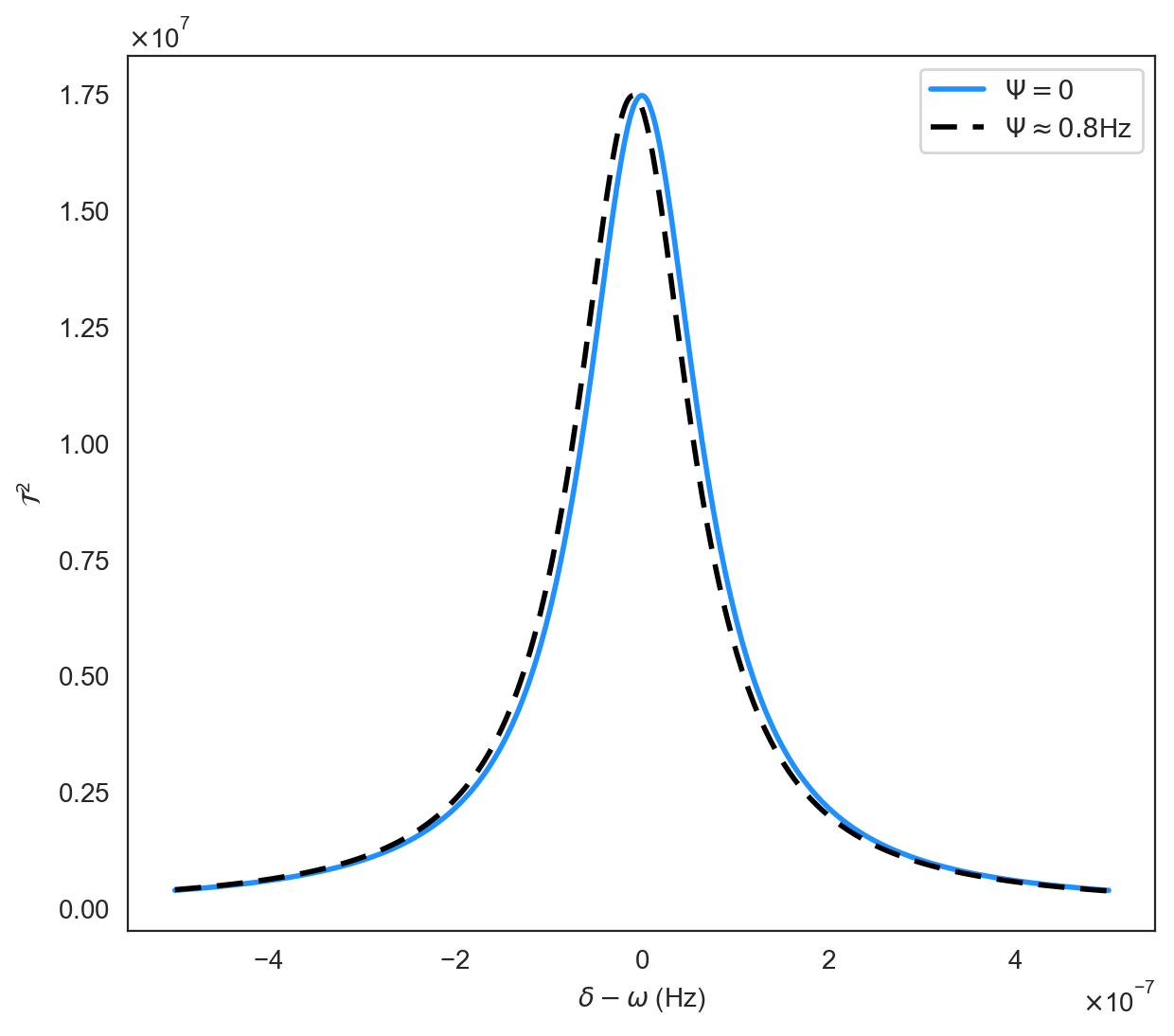}
\caption{\label{fig:3} The transmission resonance $\mathcal{T}^2$ of the probe field. The blue solid curve shows  the situation without the consideration of dipole coupling ($\Psi=0$), while the black dashed curve corresponds to the case with the coupling strength $\Psi\approx0.8\,\mathrm{Hz}$. The two spectra almost overlap, showing only a very small frequency shift. This shift is far below the minimum resolvable frequency scale of $\sim10^{-7}\,\mathrm{Hz}$ and therefore does not affect the final sensitivity or the resulting constraints. }
\end{figure}

The total Hamiltonian of the system is then given by
\begin{align}
H = {} &
\hbar \Delta_{\mathrm{pu}} \hat{a}^\dagger \hat{a}
+ \hbar \omega_a \hat{o}_a^\dagger \hat{o}_a
+ \hbar \omega_s \hat{o}_S^\dagger \hat{o}_S
+ \hbar g \hat{a}^\dagger \hat{a}(\hat{o}_a^\dagger + \hat{o}_a)
\nonumber \\
& - i\hbar \Omega_p (\hat{a} - \hat{a}^\dagger)
- i\hbar \Omega_{\mathrm{pr}}
(\hat{a} e^{i\delta t} - \hat{a}^\dagger e^{-i\delta t})
+ H_E,
\label{eq:3.1.6}
\end{align}
in which we apply the same treatment as what we did in Eqs.~\eqref{eq:2.2.1} and \eqref{eq:2.2.6}-\eqref{eq:2.2.21}. And we plot the comparison when the influence of the membrane is considered or not in Fig.~\ref{fig:3}. 

Despite the relatively large dipole coupling strength $\Psi\approx0.8\,\mathrm{Hz}$, the probe transmission spectrum shows only a negligible displacement owing to the substantial detuning between the resonance frequencies of $\omega_S$ and $\omega_a$. The induced shift is well below the minimum frequency resolution of the system. Consequently, the electromagnetic noise of membrane can be neglected, enabling a clear observation of the symmetron-induced spectral splitting.

Here now we consider the thermomechanical noise. This noise originates from the stochastic thermal motion inherent to any realistic mechanical oscillator. Its impact on the measurement is characterized by the noise spectral density $S(\omega)$ together with the effective measurement bandwidth $\Delta f$~~\cite{10.1143/PTP.49.1516}. For a one-dimensional harmonic oscillator with $\omega_n$ relevant to our system, the displacement spectral density $S_x(\omega)$ can be expressed as~~\cite{10.1063/1.1642738}

\begin{eqnarray}
S_x(\omega) = \frac{S_F(\omega )}{m^2 \big[(\omega^2 - \omega_n^2)^2 + \omega^2 \omega_n^2/Q^2\big]} .
\label{eq:3.1.7}
\end{eqnarray}

Here, the force noise spectral density is given by
$S_F(\omega) = 4 m_{eff} \omega_n k_b T / Q$, where $k_B = 1.38\times10^{-23}$J/K is the Boltzmann constant and $Q=\omega_a/2\pi\gamma_a\approx10^{12}$ the mechanical quality factor. The measurable bandwidth $\Delta f\approx10^{-5}$Hz is determined by the characteristic response time of the oscillator $\tau\approx1.6\times10^{4}$s  through the relation $\Delta f \approx 1/(2\pi \tau)$~\cite{articlejl}. 

The minimum resolvable frequency shift is obtained by integrating the spectral density of frequency fluctuations, $S_{\omega}(\omega) = (\omega_n/2Q)^2 \cdot \big(S_x(\omega)/\langle x^2_{\mathrm{rms}}\rangle\big)$, over the frequency window $\omega_n \pm \pi \Delta f$~~\cite{10.1063/1.1499745,Robins1984PhaseNI}. This procedure yields
\begin{eqnarray}
\delta\omega \approx
\sqrt{\frac{k_B T \Delta f}{m\omega_n \langle x_{\mathrm{rms}}\rangle^2 Q}} .
\label{eq:3.1.8}
\end{eqnarray}
Here, $\langle x_{\mathrm{rms}} \rangle$ denotes the root-mean-square displacement for which the pump field remains within the predominantly linear response regime. For a pump beam with a Gaussian intensity profile and beam waist $w=0.12\mathrm{mm}$, the displacement is given by $x_{\mathrm{rms}}\ll\sqrt{w^{2}/2}=84.8\mu\mathrm{m}$. Without loss of generality, we take $x_{\mathrm{rms}}=1\mu\mathrm{m}$ in the following analysis. Under this condition, the minimum detectable frequency shift limited by thermomechanical noise is estimated to be $\delta\omega_{\mathrm{min}}\approx4.32\times10^{-8}\mathrm{Hz}$, which is significantly smaller than the resonance linewidth $\omega_{\mathrm{FWHM}}\approx1.501\times10^{-7}\mathrm{Hz}$ obtained in Sec.~\ref{3.1}. Therefore, the influence of thermomechanical noise can be safely neglected.

\subsection{Constraints on the symmetron}\label{3.2}
In this section, we establish constraints on the parameters of the symmetron model based on the results discussed in Sec.~\ref{3.1}. 

Before addressing the symmetron force constraints in detail, it is necessary to determine the range of feasible values of $\mu$ imposed by the geometry of the experimental setup. The lower bound on $\mu$ is related to the cavity length $L$ along the resonant direction. In particular, the cavity mirrors on both sides should exert a negligible force on sphere A. The symmetron-induced force exerted by a single planar mass is given by~\cite{Li_2025,PhysRevD.97.064015} 
\begin{eqnarray}
F=-\frac{4\pi \phi_0^{2}\mu R}{\sqrt{2}}\tanh(\frac{\mu x}{\sqrt{2}})\text{sech}^{2}(\frac{\mu x}{\sqrt{2}}).
\label{eq:3.2.1}   
\end{eqnarray}
Here $x\approx L/2=1.55$cm represent the distance from the source plate to the sphere. Requiring the mirror-induced force to be sufficiently suppressed leads to the condition $\mu L/2 >1$, which sets the lower bound on $\mu$.  The upper bound on $\mu$ follows from the requirement $\mu d < 1$, as discussed in Eq.~\eqref{eq:2.1.14}~\cite{PhysRevD.101.064065,PhysRevD.96.124029}. 
With $d=4~\mu\mathrm{m}$, the resulting feasible range of $\mu$ is $10^{-4}$eV $< \mu < 10^{-2}$eV.

Having fixed the allowed range of $\mu$, we next establish constraints on the parameters $M$ and $\lambda$. We first consider the $M$--$\mu$ parameter plane, which is primarily determined by the background matter density.

To ensure the calculation in Sec.~\ref{2.1} and Eq.\eqref{eq:3.2.1} remain valid, the critical density $\rho_{crit}=\mu^2 M^2$ should satisfy: 
\begin{eqnarray}
\rho_{vac} < \mu^2 M^2 < \rho_{sphere},
\label{eq:3.2.2}   
\end{eqnarray}
by which the symmetron field value remains $\phi\sim0$ everywhere inside the spheres and $\phi\sim\phi_0$ in the vacuum region inside the objects in the system. Here  $\rho_{vac}\approx 2.2\times10^{-13}\text{kg}\cdot \text{m}^{-3}$ corresponds to the residual air density at $10^{-10}$torr vacuum and 200K and $\rho_{sphere}= 2.2\times10^{3}\text{kg}\cdot \text{m}^{-3}$ is the density of silica.

Additionally, the sphere are assumed to be strongly screened in Sec.\ref{2.2}, so that the thin-shell factor $\alpha>1$ is required. Considering $\alpha=\rho R^2/M^2$, the constraint can be written as 
\begin{eqnarray}
\frac{\rho_{sphere}R^2}{M^2} > 1.
\label{eq:3.2.3}   
\end{eqnarray}

\begin{figure}[h]
\centering
\includegraphics[width=0.93\linewidth]{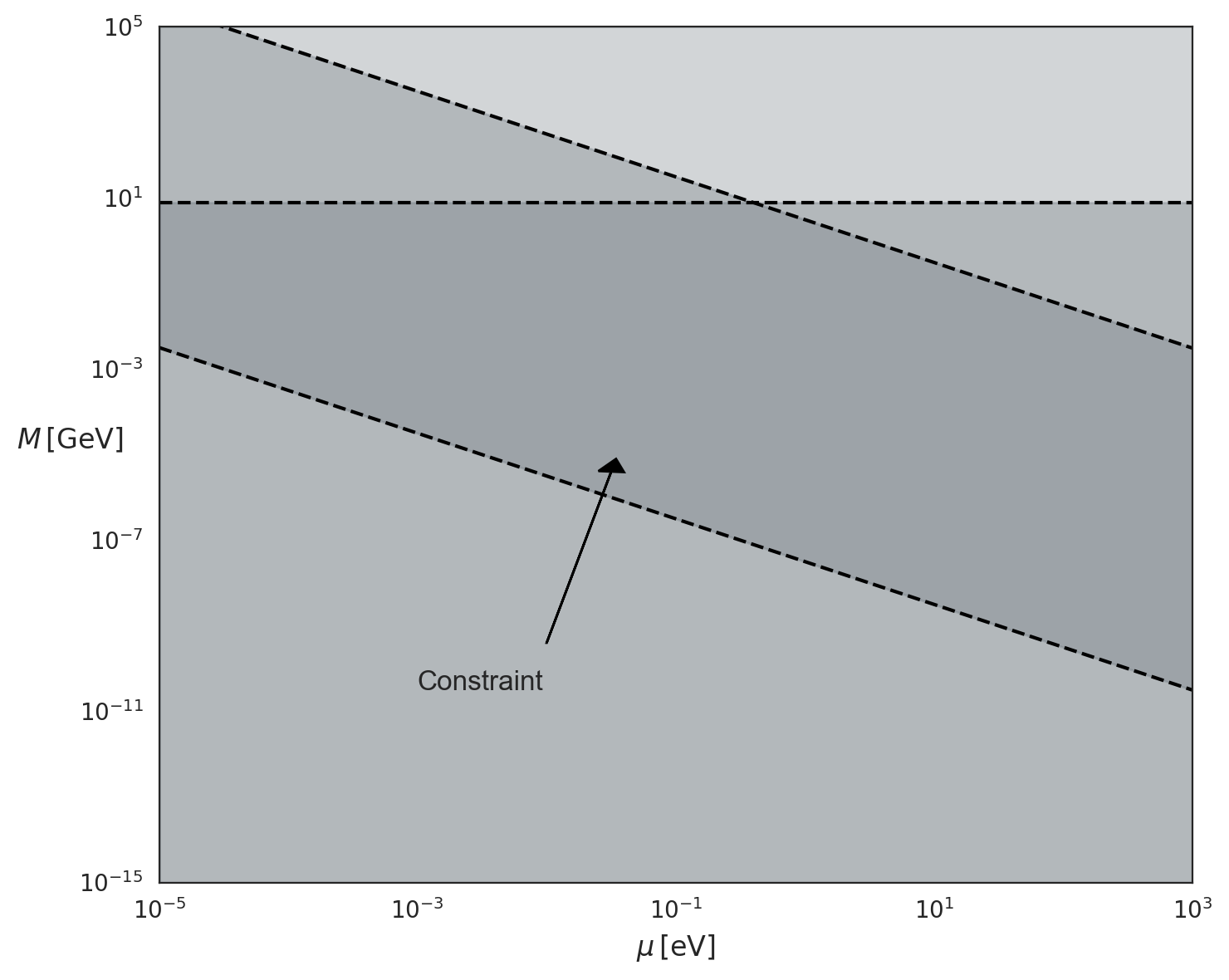}
\caption{\label{fig:4} Constraints on the symmetron parameters $M$ and $\mu$.  The shaded regions correspond to parameter spaces restricted by different theoretical and physical requirements, while the overlap region highlighted in dark grey represents the final experimental allowed parameter space.}
\end{figure}

We illustrate the resulting constraints in Fig.~\ref{fig:4} according to Eqs.~\eqref{eq:3.2.2} and \eqref{eq:3.2.3}. 
The horizontal line corresponds to the condition in Eq.~\eqref{eq:3.2.2}, while the other two lines represent the limits on $M$ imposed by Eq.~\eqref{eq:3.2.3}. 
The central region shaded in dark grey indicates the constraints obtained.

\begin{figure}[h]
\centering
\includegraphics[width=0.93\linewidth]{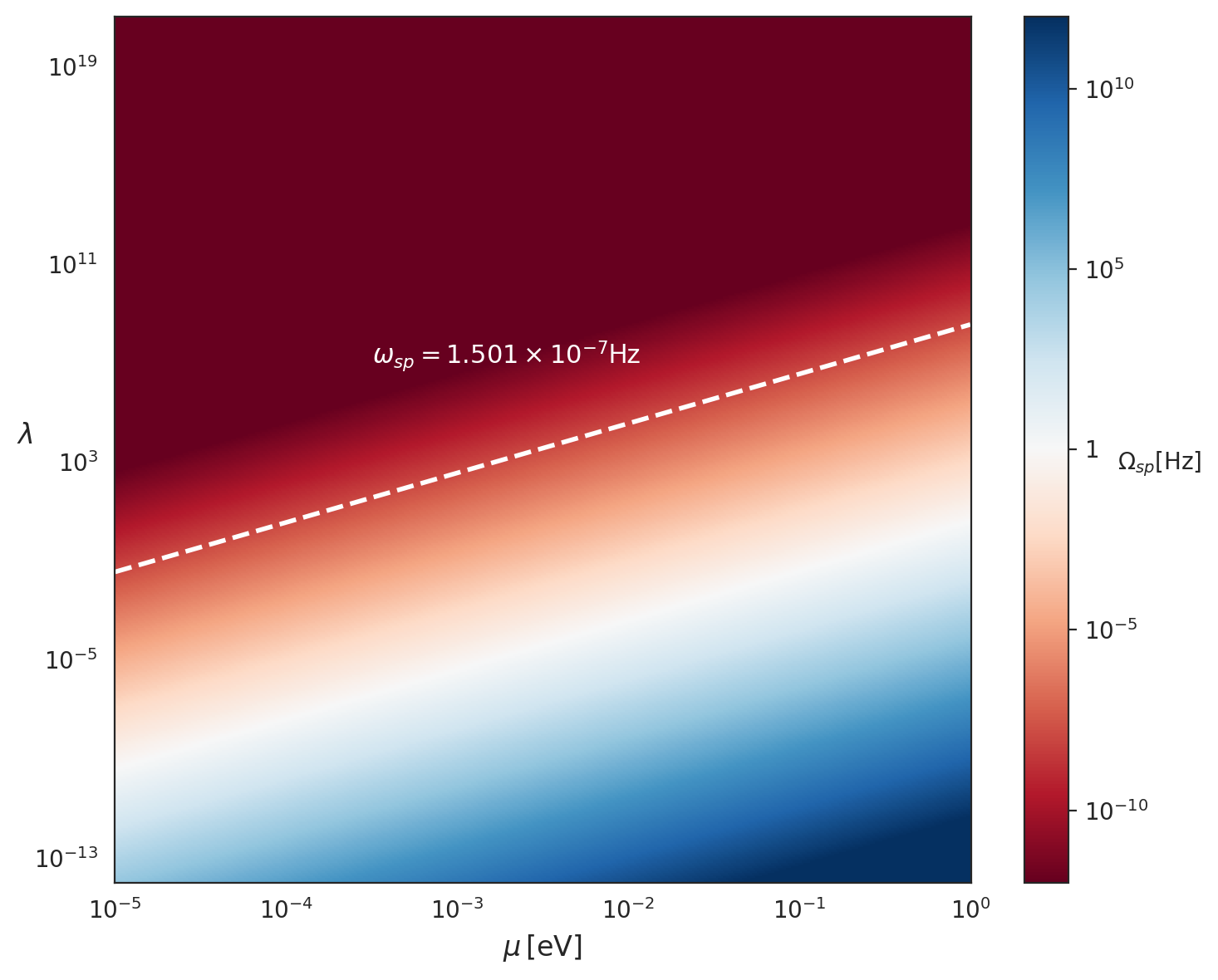}
\caption{\label{fig:5} Expected constraints on the symmetron parameters $\mu$ and $\lambda$. The color map shows the predicted frequency shift $\omega_{\rm sp}$ induced by the symmetron interaction as a function of $\mu,\lambda$. The white dashed line corresponds to the contour $\omega_{\rm sp}=1.501\times10^{-7}\,\mathrm{Hz}$, which represents the projected experimental sensitivity and thus defines the detectability threshold. Parameter regions above this curve yield splittings below the sensitivity limit and are therefore not accessible to our system. }
\end{figure}

\begin{figure}[h]
\centering
\includegraphics[width=0.93\linewidth]{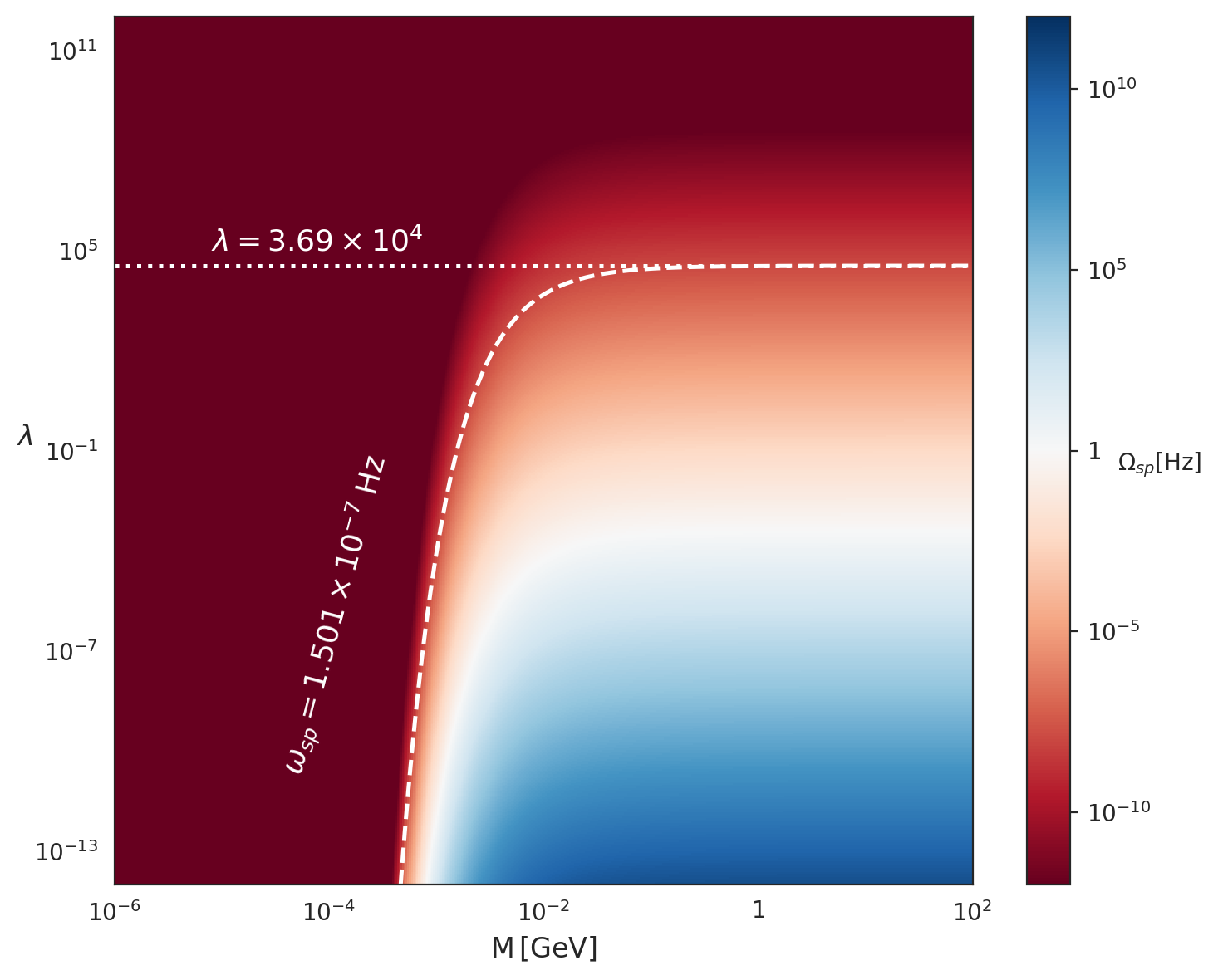}
\caption{\label{fig:6} Effect of membrane screening on the symmetron-induced peak splitting in the $\lambda$-M plane for a fixed $\mu=10^{-2}\,\mathrm{eV}$. The color map shows the predicted splitting $\omega_{\rm sp}$ including the suppression factor $e^{-2m_{\rm eff}l_m}$.The white horizontal line with label $\lambda=3.69\times10^{4}$ represents the original constraint from Fig.~\ref{fig:5}, while the white dashed curve marks the sensitivity contour $\omega_{\rm sp}=1.501\times10^{-7}\,\mathrm{Hz}$.}
\end{figure}

\begin{figure*}
\centering
\includegraphics[width=0.94\textwidth]{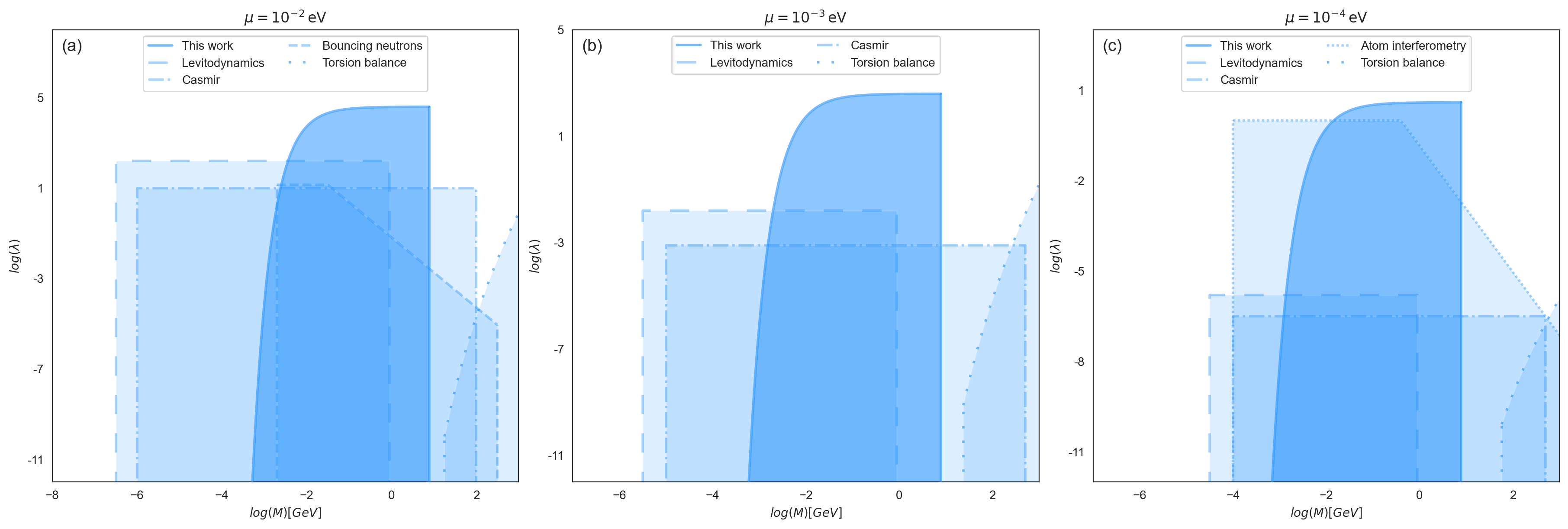}
\caption{\label{fig:7}Constraints on the symmetron parameter space from our proposed scheme, compared with bounds from Casimir-force, atom-interferometry, torsion-balance,  ultracold neutron and Levitodynamics experiments~~\cite{PhysRevD.101.064065,PhysRevLett.123.061102,PhysRevLett.110.031301,article,Li_2025}. The exclusion region is obtained from Figs.~\ref{fig:5} and \ref{fig:6} for $\mu=10^{-2}\mathrm{eV}$, $10^{-3}\mathrm{eV}$, $10^{-4}\mathrm{eV}$.}
\end{figure*}

The final step required for a complete constraint on the symmetron field is the determination of the parameter $\lambda$. The projected bound on $\lambda$ follows directly from our optomechanical sensitivity via Eqs.~\eqref{eq:2.2.5} and \eqref{eq:3.1.2-1}. For fixed values of $\mu$ and $M$, a larger peak splitting $\omega_{sp}$ corresponds to a smaller $\lambda$. Consequently, only parameter values that predict a peak splitting exceeding $\omega_{sp}^{\rm lim}=1.501\times10^{-7}\,\mathrm{Hz}$ can be probed by our scheme, while smaller splittings remain experimentally inaccessible. In Fig.~\ref{fig:5}, we illustrate the value of the peak splitting $\omega_{sp}$ with different values of $\lambda$ and $\mu$. The white line with the label $\Omega_{sp} = 1.501\times 10^{-7}\ \mathrm{Hz}$ refers to the detection limit of our system. 

Before deriving the constraints discussed in Sec.~\ref{3.2}, it is necessary to account for the suppressing effect of the membrane on the symmetron-mediated interaction acting on sphere~B. A dense membrane placed between the source and probe objects attenuates the propagation of the scalar field through the material. In the high-density environment of the membrane, where $\rho_{\rm mbr}$ is large, the symmetron field resides near the symmetry-restored configuration $\phi\simeq 0$, and small perturbations obey the following relation by dropping the high order terms 
\begin{equation}
\nabla^2\phi \simeq m_{\rm eff}^2\phi,
\qquad
m_{\rm eff}=\sqrt{\left|\frac{\rho_{\rm mbr}}{M^2}-\mu^2\right|}.
\end{equation}
Accordingly, the field amplitude across a membrane of thickness $l_m$ is exponentially attenuated, which we parameterize by a factor $t_m\sim e^{-m_{\rm eff}l_m}$ following the common treatment in torsion-balance estimates~\cite{PhysRevD.86.102003,PhysRevLett.110.031301}.

For symmetron models the fifth-force acceleration depends quadratically on the field amplitude considering Eq.~\eqref{eq:2.1.15}. So that if $\phi\rightarrow t_m\,\phi$ inside and beyond the membrane, then ${\nabla}\phi\rightarrow t_m\,{\nabla}\phi$ and the acceleration (or force signal) is suppressed as $a\rightarrow t_m^2 a$. We therefore define an effective screening factor for the symmetron-induced force,
\begin{equation}
f_{\rm scr}\equiv t_m^2 \sim e^{-2m_{\rm eff}l_m},
\end{equation}
which rescales the interaction-induced peak splitting.

Including this suppression factor, the peak splitting induced by the symmetron-mediated interaction becomes
\begin{eqnarray}
\omega_{sp}=-\frac{8\pi R^2\mu^2 f_{\rm scr}}{\lambda m_a \omega_a}.
\label{eq:3.2.4}
\end{eqnarray}
Taking $l_m\sim10^{-2}\mathrm{eV}^{-1}$ and $\rho_{\mathrm{mbr}}\sim10^{18}\mathrm{eV}^4$ in natural units, we plot the relation between $\lambda$ and $M$ for a fixed value $\mu=10^{-2}\mathrm{eV}$ in Fig.~\ref{fig:6}. A sharp cutoff appears at small $M$, corresponding to the regime in which the membrane strongly suppresses the symmetron field, while for sufficiently large $M$ the result approaches the unscreened limit shown in Fig.~\ref{fig:5}.

Fig.~\ref{fig:7} summarizes the constraints obtained for several representative values of $\mu$. The projected bounds from this work are shown in dark blue, while existing limits from other approaches are indicated by the light-blue shaded regions. These include the levitodynamics proposal~~\cite{Li_2025}, a Casimir-force–based design~~\cite{PhysRevD.101.064065}, atom interferometry experiments~~\cite{PhysRevLett.123.061102}, torsion-balance measurements~~\cite{PhysRevLett.110.031301}, and ultracold bouncing neutron experiments~~\cite{article}. As illustrated in the figure, the scheme in this work provides stronger constraints especially for $\mu$ around $10^{-3}$eV than the existing techniques, which are largely complementary to other probes. Taken together, the collection of existing constraints and projected sensitivities covers a substantial fraction of the symmetron parameter space.

\section{conclusion and discussion}\label{4}
In this work, we propose an optomechanical approach to probe and constrain the symmetron-mediated fifth force. We consider a system consisting of two high-density spheres trapped inside an optical cavity and separated by a highly reflective membrane, and analyze the resulting symmetron field configuration in the presence of screening effects. Within the framework of quantum optics, we derive the transmission spectrum of the optomechanical system and characterize the dominant noise sources relevant for force detection. This analysis establishes a quantitative connection between the symmetron-induced interaction and the minimum resolvable spectral peak splitting. Based on the projected performance of the proposed setup, we obtain constraints on the symmetron parameter space that significantly improve existing bounds, achieving an enhancement of up to four orders of magnitude around $\mu = 10^{-3}\,\mathrm{eV}$.

It is interesting to explore potential applications of our results to broader contexts, such as astrophysical phenomena including gravitational-wave detection~~\cite{fn5d-mrsj,2w9f-yy8g}. The optomechanical platform may also be relevant for next-generation experiments aimed at detecting ultraweak effects associated with dark energy~~\cite{PhysRevD.109.123023,PhysRevD.109.124007}. Compared with our previous work in Ref.~~\cite{Li_2025}, the present scheme successfully raises the upper bound on the coupling parameter $\lambda$, as shown in Fig.~\ref{fig:7}. However, the accessible ranges of $\mu$ and $M$ become narrower, mainly due to the suppressing effect introduced by the membrane. Alternative experimental designs may help to relax these constraints. Recent advances in coherent control of massive nanoparticles, including matter-wave interferometry~\cite{Gerlich2026} and quantum squeezing in levitated optomechanical systems~\cite{doi:10.1126/science.ady4652}, suggest that future extensions of our scheme could exploit similar techniques to improve coherence, increase test-mass size, and further enhance force sensitivity. In addition, employing nonspherical source or test masses could further enhance the sensitivity~~\cite{Burrage_2018,PhysRevD.91.065030}, since optimized geometries, such as ellipsoids, are expected to suppress screening effects~~\cite{PhysRevD.96.124029}. More accurate approximations for complex geometries may also lead to more reliable constraints~~\cite{PhysRevD.76.124034,PhysRevLett.98.050403}. With continued advances in experimental techniques, we expect that optomechanics-based detection schemes will play an increasingly important role in testing screened dark energy models.

\begin{acknowledgments}
This work is supported by Natural Science Foundation of Shanghai.
\end{acknowledgments}

\nocite{*}

\bibliography{apssamp}

\end{document}